\documentclass[twocolumn,english,floats,showpacs,prl]{revtex4}
\usepackage[T1]{fontenc}
\usepackage[latin9]{inputenc}
\usepackage{amsmath}
\usepackage{graphicx}
\usepackage{amssymb}

\makeatletter
\@ifundefined{textcolor}{}
{%
 \definecolor{BLACK}{gray}{0}
 \definecolor{WHITE}{gray}{1}
 \definecolor{RED}{rgb}{1,0,0}
 \definecolor{GREEN}{rgb}{0,1,0}
 \definecolor{BLUE}{rgb}{0,0,1}
 \definecolor{CYAN}{cmyk}{1,0,0,0}
 \definecolor{MAGENTA}{cmyk}{0,1,0,0}
 \definecolor{YELLOW}{cmyk}{0,0,1,0}
 }

\usepackage[above]{placeins}\usepackage{babel}

\makeatletter

\@ifundefined{textcolor}{}
{
 \definecolor{BLACK}{gray}{0}
 \definecolor{WHITE}{gray}{1}
 \definecolor{RED}{rgb}{1,0,0}
 \definecolor{GREEN}{rgb}{0,1,0}
 \definecolor{BLUE}{rgb}{0,0,1}
 \definecolor{CYAN}{cmyk}{1,0,0,0}
 \definecolor{MAGENTA}{cmyk}{0,1,0,0}
 \definecolor{YELLOW}{cmyk}{0,0,1,0}
 }
\makeatother

\makeatother

\usepackage{babel}

\makeatother

\usepackage{babel}

\makeatother

\usepackage{babel}

\makeatother

\usepackage{babel}

\makeatother

\usepackage{babel}

\begin{document}

\title{Are the high-$T_{c}$ superconductors strongly correlated electron
systems?}

\author{X. Q. Huang$^{1,2}$}

\email{xqhuang@netra.nju.edu.cn}

\affiliation{$^{1}$Department of Telecommunications Engineering ICE, PLAUST,
Nanjing 210016, China \\
 $^{2}$ Department of Physics and National Laboratory of Solid
State Microstructure, Nanjing University, Nanjing 210093, China}

\date{\today}
\begin{abstract}
In this paper, we argue that the high-temperature superconductors
do not belong to strong correlated electron systems. It is shown that
both the two-dimensional Hubbard and \emph{t}-\emph{J} models are
inadequate for describing high temperature superconductivity. In our
opinion, a superconducting phase should be an energy minimum electronic
state which can be described in a new framework where the electron-electron
interactions (both on-site Hubbard term and off-site term) and the
electron-phonon interaction can be completely suppressed.
\end{abstract}


\maketitle
The mechanism of high temperature superconductivity, despite great
efforts from both theoretical and experimental approaches since 1986\cite{bednorz},
remains an almost complete mystery in condensed matter physics. As
is well known, the undoped parent compounds of the high temperature
superconductors are so-called Mott insulators and the superconductivity
can be achieved by doping carriers into these materials. It is widely
believed in the physical community that a comprehensive understanding
of the doping evolution from insulating to superconducting state may
help to uncover the underlying mechanisms of high temperature superconductivity.
However, although under intensive studies for about twenty-five years,
physicists do not agree on how superconductivity works in these materials.

Theoretically, it is generally accepted that the conventional superconductors
are well-described by the BCS pairing theory of the electron-phonon
interaction mechanism\cite{BCS}, while the high-temperature superconductors
are strongly correlated electron systems that may need a new (or at
least improved) theory. Many researchers blindly believe that the
Hubbard\cite{Hubbard} and \emph{t}-\emph{J} models\cite{t-J} can
capture the essential physics of the high-temperature superconducting
phenomenon. Thousands of articles based on these two models have been
published during the past several decades. Unfortunately, no one of
these works is now known to be a valid interpretation of the phenomena
in cuprate superconductors. We think that in such a situation researchers
should consider some fundamental issues of high-temperature superconductivity,
for example, do the high-temperature superconductors belong to strong
correlated electron systems? Are the Hubbard and \emph{t}-\emph{J}
models valid to be used to describe the superconductivity?

In this paper, we argue that the old theoretical framework of the
superconductivity has an unrecognized flaw. Based on the energy minimum
principle and the experimental observation of quasi-one-dimensional
charge stripes\cite{tranquada}, a new framework is proposed to describe
all superconductors.

\begin{figure}
\begin{centering}
\resizebox{1\columnwidth}{!}{ \includegraphics{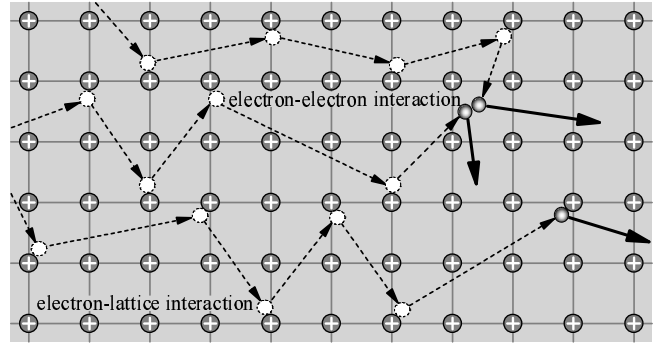}}
\par\end{centering}

\caption{In classical physics, the resistance mechanism consists of two parts:
the main part is contributed by the electron-lattice interactions
(scattering), another part is a result of electron-electron interactions
(collision), as illustrated in this figure. In our opinion, the most
important question concerning the mechanism of superconductivity will
be converted into a simple question: how can these two kinds of interactions
be avoided when the material goes into the superconducting state?}

\label{fig1}
\end{figure}

Let's start with a very basic question: What causes the resistance
in the metal materials? In the framework of classical physics, a metal
consists of a lattice of positive ions surrounded by a sea of electrons
that will drift from one end of the metal to the other under the influence
of an applied electric field. During this procedure, the electrons
will lose some of their kinetic energy due to the scattering by the
thermal motion of ions or collision with other electrons, as shown
in Fig. \ref{fig1}, which are the two main source of resistance.
Obviously, when a material is driven into the superconducting state,
the superconducting electrons will no longer be scattered by the ions,
and the electron-electron collisions are also completely eliminated
in the superconductor.

\begin{figure*}
\begin{centering}
\resizebox{1.8\columnwidth}{!}{ \includegraphics{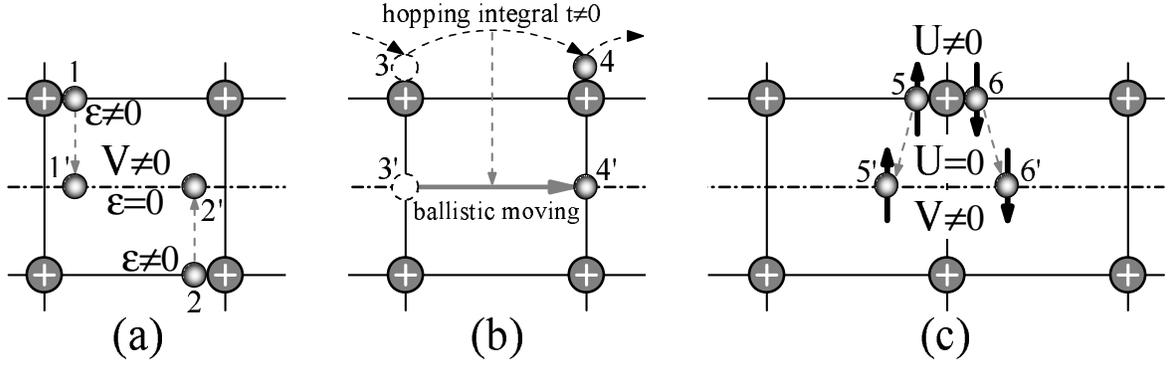}}
\par\end{centering}

\caption{The explanation of why the three physical parameters (on-site energy
$\varepsilon$, hopping interaction $t$ and the strong Hubbard interaction
$U$) given in Hamiltonian form in second-quantization theory are
invalid to be used for describing the phenomenon of superconductivity.
(a) The on-site energy $\varepsilon$, (b) hopping interaction $t$,
and (c) the Hubbard interaction $U$, they can very easily be excluded
from the Hamiltonian if electrons are located in the equilibrium positions
(1' , 2', 3', 4', 5' and 6') of zero potential energy.}

\label{fig2}
\end{figure*}

To be a reliable theory of superconductivity, it must provide a clear
description or explanation of how the two kinds of interactions (see
Fig. \ref{fig1}) can be fully suppressed when a superconductor goes
from the normal state into the superconducting state. Unfortunately,
almost all existing theories of superconductivity have neglected these
basic physical facts. So far, the researchers in the field of superconductivity
are used to do their calculations by blindly applying the Hamiltonian
handed down from generation to generation, physically, it is most
likely that they are on the wrong track. Note that the complex numerical
simulations cannot generate any results that are not already implied
in the framework. In our opinion, the reason why the mechanism of
high temperature superconductivity still remains unknown lies in the
fact that the key factors of Fig. \ref{fig1} are not considered in
the Hubbard and \emph{t}-\emph{J} models.

The Hubbard model is the simplest model of interacting particles in
a lattice, which is mathematically defined by following Hamiltonian
\begin{equation}
H=-t\sum_{\left\langle ij\right\rangle \sigma}(c_{i\sigma}^{\dagger}c_{j\sigma}+c_{j\sigma}^{\dagger}c_{i\sigma})+U\sum_{i}n_{i\uparrow}n_{i\downarrow}\text{,}\label{hb}\end{equation}
 where $t$ is the hopping matrix element between the nearest neighbor
sites of the lattice, $U$ is the on-site Coulomb repulsion, $c_{i\sigma}$
($c_{i\sigma}^{\dagger}$) annihilates (creates) an electron with
spin $\sigma$ at site $i$, and $n_{i\sigma}=c_{i\sigma}^{\dagger}c_{i\sigma}$.

In 1977, Jozef Spalek derived the so-called \emph{t}-\emph{J} model
from the above Hubbard model. The corresponding Hamiltonian also consists
of two pieces:

\begin{eqnarray}
H & = & -t\sum_{\left\langle ij\right\rangle \sigma}\left[c_{i\sigma}^{\dagger}(1-n_{i-\sigma})(1-n_{j-\sigma})c_{j\sigma}+H.c.\right]\notag\\
 &  & +J\sum_{\left\langle ij\right\rangle }(S_{i}S_{j}-\frac{1}{4}n_{i}n_{j}).\label{tj}\end{eqnarray}
 where $t$ is an effective transfer integral, $J$ ($=4t^{2}/U$)
is the antiferromagnetic exchange energy for a pair of nearest neighbor
sites $\left\langle ij\right\rangle $, $S_{i}$ and $S_{j}$ are
spin-1/2 operators.

\begin{figure*}
\begin{centering}
\resizebox{1.6\columnwidth}{!}{ \includegraphics{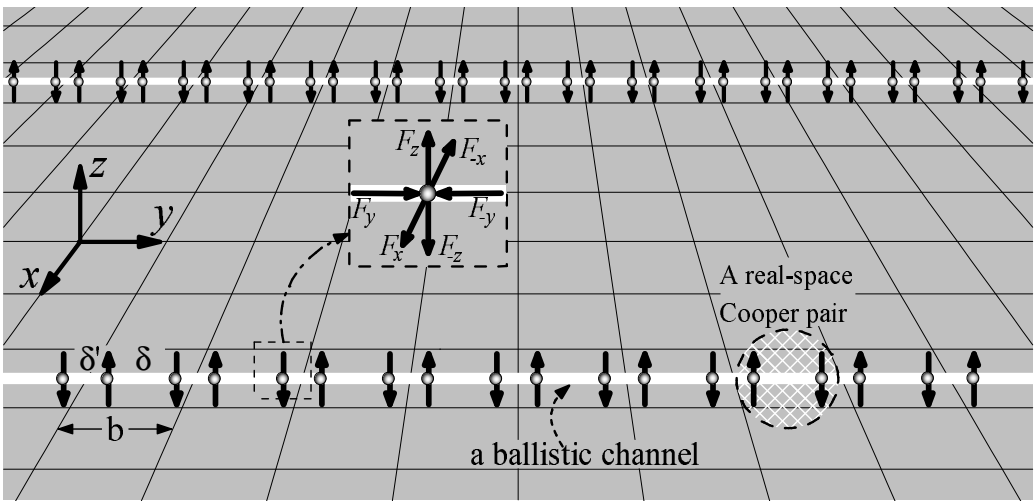}}
\par\end{centering}

\caption{A real-space superconducting ground state where the charge carriers
(electrons) can self-assemble into some static one-dimensional peierls
charge and spin antiferromagnetic stripes. In this special case, it
is easy to prove that the resultant force applied to each electron
is equal to zero along any directions ($\left|F_{x}\right|=\left|F_{-x}\right|$,
$\left|F_{y}\right|=\left|F_{-y}\right|$ and $\left|F_{z}\right|=\left|F_{-z}\right|$).
In other words, the strong electron-electron Coulomb interactions
within the stripe can be completely suppressed under this simple framework
($\left|F_{y}\right|=\left|F_{-y}\right|$) . Under the influence
of the external fields, the peierls chains will transfer into periodic
chains and the superconducting current can flow without resistance
along the ballistic channels.}

\label{fig3}
\end{figure*}

Do the Hubbard and \emph{t}-\emph{J} models capture the key physics
of the high-temperature superconductivity?

From the energy point of view, a stable superconducting state must
first be an energy minimum electron coherent state with the complete
suppression of the electron-electron and electron-ion interactions
inside the superconductor. However, the theoretical frameworks of
Hubbard and \emph{t}-\emph{J} models apparently overlook these factors.

According to the Hamiltonian of Eq. (\ref{hb}), if two electrons
are on the same site (with oppositely directed spin), the second term
will be extremely large because of strong on-site Coulomb repulsion.
Although this hypothesis does not violate the Pauli exclusion principle,
the configuration with two electrons of opposite spin on the same
site is harmful to the formation of the superconducting state. Hence,
the second term of Eq. (\ref{hb}) must be ruled out. Without
the contribution of the second term, the Hubbard model will automatically
change into the conventional tight binding model from regular band
theory. Of course, the retained first term of Eq. (\ref{hb}) is also
impossible to describe the superconducting behavior, since the hopping
picture is in fact an idealized model of electron-ion interactions
(scattering) as shown in Fig. \ref{fig1}. In other words, both the
two terms of the Hubbard Hamiltonian are contrary to the basic assumption
of superconductivity that the electron-electron and electron-ion interactions
should be completely eliminated. We expect that all the above discussions
are also suitable for the \emph{t}-\emph{J} model of Eq. (\ref{tj}).

Obviously, the Hubbard and \emph{t}-\emph{J} models based on the Hamiltonian
cannot be applied to explain superconductivity. In the following,
we will discuss how to establish a correct theory of superconductivity.
In the Wannier representation, the Hamiltonian can generally be decomposed
into three parts, as shown in Fig. \ref{fig2}. The first part is
the so-called on-site energy $\varepsilon$ which originates from
the strong short-range attraction between electron and ions, as marked
by 1 and 2 in Fig. \ref{fig2}(a). This strong interaction ($\varepsilon\neq0$
) can be converted quite easily to the weak long-range electron-electron
interaction ($V\neq0$, $\varepsilon=0$), if the two electrons are
located in the equilibrium positions {[}1' and 2' in Fig. \ref{fig2}(a){]}
of zero potential energy. The second part is the hopping interaction
$t$ which is an extremely simplified picture (without involving any
detailed hopping mechanism) of kinetic electron-ion interaction {[}from
3 to 4 in Fig. \ref{fig2}(b){]}. As discussed above, the hopping
picture is unsuitable to describe the phenomenon of the superconductivity.
A more reasonable picture is to assume that the electron can tunnel
from 3' to 4' through a ballistic channel, as also shown in Fig. \ref{fig2}(b).
The third part is the strong on-site electron repulsive interaction
(5 and 6 in Fig. \ref{fig2}(c) with Hubbard $U\neq0$) which can
also be easily avoided if the two electrons are assumed to be arranged
in the equilibrium positions 5' and 6', as also shown in Fig. \ref{fig2}(c).
In this case, the strong Hubbard interaction (elephant $U$) is replaced
by a ``soft'' electron-electron interaction (mouse $V$).

Based on the minimum energy principle, all the charge carriers have
the tendency to stay around the equilibrium positions with only the
long-range (compared to the on-site Hubbard interaction) electron-electron
Coulomb repulsion $V\neq0$. Hence, the microscopic mechanism of the
high-temperature superconductivity turns out to be a very simple problem:
How can the repulsive interaction $V$ between electrons be overcome
in favor of the superconductivity? Obviously, the strong direct Coulomb
repulsion between electrons cannot be suppressed by exchanging a small
``second-order'' quasiparticle, for example, the quantized phonon
induced by the lattice vibration. In fact, the strongly repulsive
electrons can be ``glued'' together purely by a real-space electronic
mechanism.

The determination of the superconducting ground state is the key theoretical
issue in the study of the mechanism of the superconductivity. What
is the superconducting ground state and how to describe it? In the
previous series papers\cite{huang1,huang2,huang3,huang4,huang5,huang6,huang7},
it has been elucidated that the formation of the quasi-one-dimensional
charge stripes in the superconductors plays the fundamental role of
the superconductivity. In our approach, the most basic unit of the
superconducting ground state is the static one-dimensional peierls
charge and spin antiferromagnetic chain\cite{huang2}, as shown in
Fig. \ref{fig3}. It has been proven analytically that without applying
the external field, the charge carriers (electrons) can self-organize
into some one-dimensional peierls charge chains with $\delta+\delta'=b$
(see Fig. \ref{fig3}), where $b$ is the lattice constant in the
stripe direction. In this situation, the resultant force applied to
each electron is exactly zero along any directions ($\left|F_{x}\right|=\left|F_{-x}\right|$,
$\left|F_{y}\right|=\left|F_{-y}\right|$ and $\left|F_{z}\right|=\left|F_{-z}\right|$)
. Moreover, a stable real-space Cooper pair can be defined inside
one plaquette, as also shown in Fig. \ref{fig3}.

Under the influence of the external fields, the transition from the
ground state into the excited state will occur for the superconducting
electrons. Consequently, the peierls chains will transfer into some
real-space periodic charge stripes with a definite electron-electron
spacing of $\delta=b/2$. For an excited superconducting state, all
electrons are identical and the electron-electron Coulomb repulsions
can be naturally suppressed due to the symmetry of the real-space
charge stripes. In this case, all the superconducting electrons will
be condensed into a coherent state and the concept of Cooper pair
will no longer has any physical meaning. In this scenario, the superconducting
current can flow without resistance along many ballistic channels,
as shown in Fig. \ref{fig3}.

Finally, we present a brief discuss about the validity of the BCS
pairing theory. Recently, Hirsch argued that it is time to question
the validity of the BCS theory of superconductivity\cite{Hirsch}.
Although we don't completely agree with all of his views, some of
his arguments are physically reasonable and interesting. In our viewpoint,
the pairing and superconductivity are two different and unrelated
physical phenomena. The electron pairing is an effect of the short-range
electron correlation, while the superconductivity comes from the long-range
electron correlation (as shown in Fig.\ref{fig3}). In fact, the pairing
mechanism cannot effectively inhibit the generation of the resistance
in the superconductor. Let us look at Fig. \ref{fig1} again, the
electron-lattice (or pair-lattice) interactions are avoidable even
if all electrons are paired. In addition, the much stronger pair-pair
Coulomb repulsion will inevitably lead to the generation of resistance.
In our opinion, it is most likely that there exists some flaws in
the electron-phonon interaction based BCS theory\cite{huang8}.

In this short letter, it has been pointed out that the high-temperature
superconductivity has nothing to do with the strong correlated electron
systems. It should be emphasized that any kind of materials materials
will no longer be the strongly correlated systems when the materials
enter into the superconducting states. We have shown clearly that
both the Hubbard and \emph{t}-\emph{J} models are unsuitable to describe
the high temperature superconductivity. It has been stressed that
a reliable theory of superconductivity should be based on the energy
minimum principle and both the electron-electron (on-site and off-site)
and electron-ion interactions must be effectively suppressed. We have
outlined a new theory of superconductivity proposed to apply to all
superconductors (both conventional and unconventional).

\end{document}